%% file: main.tex
\documentclass[twocolumn,aps,floatfix,preprintnumbers,showpacs,amsfonts,amssymb,prb,superscriptaddress,showkeys]{revtex4}
\usepackage{graphicx,epsf}
\usepackage[T1]{fontenc}
\usepackage{multirow}
\usepackage[latin1]{inputenc}
\usepackage{amsmath}
\usepackage{revsymb}
\usepackage{amssymb}
\usepackage{bm}

\makeatletter

\usepackage{amscd}
\usepackage{bm}
\makeatother

\input{newcommands.tex}

\newcommand{\LDOSsize}{1.6cm}
\newcommand{\LDOSsizeU}{1.1cm}
\newcommand{\DOSsize}{2.5cm}
\newcommand{\NRsize}{2.22cm}
\newcommand{\NRBARsize}{2.93cm}

\begin{document}

\title{Local density of states of electron-crystal phases in graphene in the quantum Hall regime}

\author{O.\ Poplavskyy}
\email[Emails:\ ]{apopl@ukr.net, op226@cam.ac.uk}
\affiliation{\affCU}
\affiliation{\affLPS}
\affiliation{\affUU}
\author{M.\ O.\ Goerbig}
\affiliation{\affLPS}
\author{C.\ Morais\ Smith}
\affiliation{\affUU}

\date{\today{}}

\begin{abstract}
We calculate, within a self-consistent Hartree-Fock approximation, the local density of states for different electron crystals in graphene subject to a strong magnetic field. We investigate both the Wigner crystal and bubble crystals with $M_e$ electrons per lattice site. The total density of states consists of several pronounced peaks, the number of which in the negative energy range coincides with the number of electrons $M_e$ per lattice site, as for the case of electron-solid phases in the conventional two-dimensional electron gas. Analyzing the local density of states at the peak energies, we find particular scaling properties of the density patterns if one fixes the ratio $\nu_N/M_e$ between the filling factor $\nu_N$ of the last partially filled Landau level and the number of electrons per bubble. Although the total density profile depends explicitly on $M_e$, the local density of states of the lowest peaks turns out to be identical regardless the number of electrons $M_e$. Whereas these electron-solid phases are reminiscent of those expected in the conventional two-dimensional electron gas in GaAs heterostructures in the quantum Hall regime, the local density of states and the scaling relations we highlight in this paper may be, in graphene, directly measured by spectroscopic means, such as e.g. scanning tunneling microscopy.
\end{abstract}


\maketitle

\section{Introduction}
\label{label.introduction.}
\indent As was shown by Wigner in 1934,\cite{wigner} the degenerate Fermi gas is unstable towards the formation of a periodic triangular lattice of localized electrons (electron crystal),\ once the Coulomb energy prevails over the kinetic one. However, the critical electron density at which the transition to the Wigner crystal (WC) occurs is too low for usual metals.  Nevertheless, the situation is much improved if one applies a strong perpendicular magnetic field to a two-dimensional (2D) electron gas (2DEG). In this case,\ the single-particle continuous energy spectrum is quantized into a sequence of hugely degenerate Landau levels (LL's). If one restricts oneself to the electrons within the last partially-filled LL, one finds that their kinetic energy is quenched, and the only energy scale is the Coulomb energy, which favors the formation of an electron crystal at small filling factors.\cite{FPA,yoshioka.fukuyama,yoshioka.lee}
\newline\indent  
A quantum electron crystal in the presence of a disorder potential is expected to become collectively pinned and to manifest itself as an insulator.\cite{FL}
While at small filling factors the 2D WC with triangular lattice symmetry\cite{yoshioka.fukuyama,yoshioka.lee} is expected to yield the global energy minimum,\ it was
predicted that the phase diagram of the 2DEG includes also electron-bubble crystals (a periodic lattice with more than one electron per site),\ stripes,\cite{fogler,morais,cote.bubbles}\ and even more exotic quantum Hall liquid-crystal phases.\cite{fogler.QHLC} 
Unlike electron-crystal phases, the prominent quantum liquids, which display the fractional quantum Hall effect in the two lowest LLs,\cite{FQHErev} are translationally and rotationally invariant and remain conducting even in the presence of disorder. 
Therefore, these different quantum phases may be distinguished experimentally with respect to the behavior in transport measurements. 
For instance, a succession of insulating and conducting phases yields a re-entrant integer quantum Hall effect (IQHE) in the first\cite{eis} ($N$=1) and second\cite{cooper} ($N=2$) excited LLs and has been interpreted in terms of a competition of such
electron-solid and quantum-liquid phases.\cite{GLM}
Further evidence for electron crystals in LLs stems from radio-frequency spectroscopy\cite{fertig.WC} and transport measurements
under microwave irradiation,\cite{lewis1,lewis2,lewis3,chen.LLLWC,chen.melting} which excites the collective pinning mode of the electron crystals.
\newline\indent While all these experimental techniques have been very successful in discovering new insulating phases and have confirmed a number of theoretical predictions, they are indirect evidence for high-field electron crystals based on transport measurements 
-- the 2DEG in GaAs heterostructures is buried deep inside the substrate, which renders impossible a direct optical observation of a periodic electron-crystal lattice, e.g., by means of scanning tunneling microscopy (STM).\cite{fischer.STM} 
In contrast to the conventional 2DEG, such optical studies might become possible in graphene, a one-atom thick sheet of graphite, with unique electronic and mechanical properties.\cite{geim.novoselov,antonioRev} Indeed, graphene may be viewed as a particular 2DEG, where
the electrons behave as if they were massless particles described by the relativistic 2D Dirac equation.
In this Dirac equation the Fermi velocity $v_F$ plays the role of the speed of light $c$, although it is roughly 300 times smaller than the latter in vacuum. In addition, the Brillouin zone of graphene
has two non-equivalent corner points (called Dirac points) which yield a twofold valley degeneracy and which may formally be described in terms of an SU(2) pseudospin degree of freedom.

In strong magnetic fields, the energy of Dirac fermions in graphene is quantized into LL's, the structure of which is different from
that of non-relativistic electrons in a conventional 2DEG. Apart from their unconventional (square root) magnetic-field dependence, there 
exists a LL at exactly zero energy, and each LL in the conduction band has a counterpart in the valence band. This particular LL structure
leads to the anomalous integral quantum Hall effect observed in graphene.\cite{novoselov,zhang} In addition, the energy gap between the subsequent LL's in graphene is so large, that it is possible to observe the IQHE even at room temperatures.\cite{roomT}

As the mobility of graphene samples is further improved, one may expect to observe the fractional quantum Hall effect, which has been studied theoretically by several authors,\cite{goerbig.interactions.graphene,AC,toke,khves,GR}
and also collectively-pinned insulating phases such as Wigner-crystal and bubble phases, as predicted in Refs. \onlinecite{zhang.joglekar,zhang.joglekar.LLM,jianhui.wang,cote.skyrme.wc.graphene,hao.wang}. 
In contrast to GaAs heterostructures, these electronic phases occur at the surface of the graphene sheet and are, thus, directly accessible by spectroscopic means. Indeed STM has been applied successfully to probe the density distribution in exfoliated\cite{martin} and epitaxial\cite{mallet} graphene, as well as in graphene on a graphite substrate in a strong magnetic field.\cite{eva}
This exciting prospect motivated us to calculate theoretically  physical quantities of a 2D electron crystal which might be measured in an STM experiment: the (integrated) density of states (DOS) and the local density of states (LDOS). We should note that the quantum Hall regime is the only case for which one may expect the formation of electron-crystal phases in graphene. Indeed, it is predicted that the 2D Wigner crystallization is completely absent in graphene for any electron density in the absence of a magnetic field,\cite{no.wc.graphene} due to the scale invariance of the dimensionless interaction parameter $r_s=e^2/\hbar\epsilon v_F\simeq 2/\epsilon$ for a 2D system with a linear dispersion relation. Here, $\epsilon$ is the dielectric constant which depends on the environment where the graphene sheet is embedded.

In this paper, we discuss the DOS and the LDOS for several electron crystals in the $N=2$ LL within a Hartree-Fock approximation. 
We have performed similar calculations for $N=1$, 3 and 4, but we concentrate in the present paper on $N=2$ for two reasons. 
First, the DOS and LDOS results for $N=2$ are representative of high-field electron solids -- our calculations yield indeed similar
results for the other LL's. Second, for higher LL's there have been no clear indications so far for electron-crystal phases in GaAs in the quantum Hall regime.
For our numerical calculations, we have adopted the iterative scheme proposed by C\^ot\'e and MacDonald,\cite{cote.macdonald,cote.brey.macdonald} which has also been applied to calculate the energies and the real-space profiles of various electron-crystal phases in graphene.\cite{zhang.joglekar} As a test of the validity of our code, we have corroborated the results obtained in Ref. \onlinecite{zhang.joglekar} and then applied it for the calculation of the DOS and the LDOS. Notice that, despite the huge amount of Hartree-Fock studies of quantum Hall electron-crystal phases, none is devoted to study the LDOS of these phases. We have calculated the LDOS at energies where the integrated DOS has well-pronounced peaks, which fall into two distinct classes: bound states at negative energy with respect to the chemical potential and high-energy peaks above. The number of negative-energy peaks is identical with the number of electrons $\Me$ per bubble, in agreement with bubble crystals in the conventional 2DEG.\cite{cote.bubbles} Furthermore, we find that the sum of the LDOS at these $\Me$ negative-energy peaks reproduces the real-space density profile of the $\Me$-electron bubble crystal. 

This paper is organized as follows. In Section \ref{sec.HF.}, we outline the basic steps of the Hartree-Fock approximation to the 2DEG in graphene.\ In Section \ref{sec.results}, we present numerical results for the DOS and the LDOS in graphene. Finally, we draw our conclusions in Section \ref{sec.concl.}.

\section{Hartree-Fock Hamiltonian}
\label{sec.HF.}

For a partially filled LL $N$, the low-energy electronic properties are captured within a model that takes into account states only within this level. In this case, the single-particle kinetic energy is the same for all of states, and thus only the interaction term is relevant. Furthermore, we omit the physical spin which we consider to be completely polarized, e.g. due to a sufficiently large Zeeman effect. The derivation of the Hartree-Fock Hamiltonian for the 2DEG in GaAs has been extensively discussed in the literature.\cite{cote.macdonald,cote.brey.macdonald} In graphene, the interaction Hamiltonian for the 2DEG is similar to that in GaAs, albeit with different form factors due to the spinorial form of the wave functions.\cite{goerbig.interactions.graphene,nomura.macdonald} This similarity allows one to use the same theoretical methods which were used previously to study the 2DEG in GaAs, with the important difference that we need to take into account the twofold valley degeneracy in the form of an SU(2) pseudospin degree of freedom, 
$\beta=\pm 1$.
Provided that inter-LL transitions are neglected, we may write the interaction part of the full Hamiltonian for the 2DEG of spinless electrons in graphene as\cite{goerbig.interactions.graphene}
\begin{align}
\label{hint}
\oHint=\frac{\Vc}{2}\sum_{\idf,\idf',\qq}\frac{1}{|\qq|}[\Fn{N}{\qq}]^2
\oiia{\rho}{\idf}{\idf}{-\qq}\oiia{\rho}{\idf'}{\idf'}{\qq},
\end{align}
where $\Vc\equiv e^2/\ML\epsilon$ is the Coulomb energy scale, with $\ML=\sqrt{\hbar/eB}$ the magnetic length, $B$ the magnetic field, and $\epsilon$ the dielectric susceptibility of the medium, and $\qq\equiv{(q_x,q_y)}$ is a 2D wavevector.
The (guiding center) density operator in the Landau gauge reads
\begin{align}
\label{densop}
\nonumber \oiia{\rho}{\idf_1}{\idf_2}{\qq}& =\invg\sum_{X}\exp\left(-iq_xX-i{\frac{\MLS q_xq_y}{2}}\right)
\\
&\quad\times\oiid{c}{X}{\idf_1}\oii{c}{X+\MLS q_y}{\idf_2}. 
\end{align}
Here, $\LLdeg=S/2\pi\MLS$ measures the LL degeneracy, with the square area $S$ of the 2DEG sample, $\oii{c}{X}{\beta}$ and $\oiid{c}{X}{\beta}$ are the electron's destruction and creation operators, respectively,\ where $X$ denotes single-particle quantum states within the $N$-th LL.
Finally, in Eq.\ \eqref{hint} the graphene form factor $\Fn{N}{\qq}$ reads \cite{nomura.macdonald,goerbig.interactions.graphene}
\begin{align}
\label{ff.graphene}
 \Fn{N}{\qq}=
 \begin{cases}
 \frac{1}{2}\left[L_{|N|}\left(\frac{q^2}{2}\right)+L_{|N|-1}\left(\frac{q^2}{2}\right)\right]e^{-\frac{q^2}{4}},&\ N\neq{0};\\
 e^{-\frac{q^2}{4}},&\ N=0, 
\end{cases}
\end{align}
where $q\equiv{|\qq|}$, and $L_n(x)$ is the Laguerre polynomial of order $n$.
We note that the 2DEG form factor in GaAs is given by\cite{cote.macdonald}
\begin{align}
\label{ff.gaas}
 F_N(\qq)=L_N\left(\frac{q^2}{2}\right)e^{-q^2/4}.
\end{align}
It is apparent from Eqs.\ \eqref{ff.graphene}\ and\ \eqref{ff.gaas} that the graphene (relativistic 2DEG) form factor is simply a linear combination of form factors for adjacent LL's of the non-relativistic 2DEG in GaAs. This peculiar fact results from mixing the Dirac particle wavefunctions between the sites of two sublattices in graphene, and is also a consequence of the spinorial nature of these wavefunctions. Apart from the difference in form factors given by Eqs.\ \eqref{ff.graphene} and \eqref{ff.gaas}, the 2DEG in GaAs and graphene is described equivalently, as follows from the same analytical structure of the Coulomb interaction term given by Eq.\ \eqref{hint}.  

Finally, we note that the Hamiltonian in Eq.\ \eqref{hint} is SU(2)-invariant with respect to the valley pseudospin. In contrast to the physical electron spin, this SU(2) symmetry is approximate. However, SU(2)-symmetry breaking terms are suppressed linearly in 
$a/l_B\ll 1$ where $a=0.14$nm is the carbon-carbon distance in graphene and $l_B=26/\sqrt{B\rm [T]}$nm, i.e. at an energy scale that is well below the disorder broadening of the LL's.\cite{goerbig.interactions.graphene,AF}
This physical model is similar to another two-component quantum Hall system -- \ if one replaces in Eq.\ \eqref{hint} $\Fn{N}{\qq}$ by the non-relativistic form-factor $F_N(\qq)$, one obtains the Hamiltonian for the non-relativistic 2DEG including the electrons' spin in the absence of a polarizing Zeeman effect. Alternatively, this model may describe a
quantum Hall bilayer in the theoretical limit of zero layer separation, where the two ``spin'' orientations denote the two different layers.\cite{cote.brey.macdonald} One may further simplify the model in Eq.\ \eqref{hint} by omitting the valley pseudospin degree of freedom, in which case one presupposes a complete valley polarization of the electronic phases, which would maximally profit from the exchange interaction. This effective U(1) model is described by the interaction term
\begin{align}
\label{u1graphene}
\oHint=\frac{\Vc}{2}\sum_{\qq}\frac{1}{|\qq|}[\Fn{N}{\qq}]^2
\oa{\rho}{-\qq}\oa{\rho}{\qq},
\end{align}
where the density operator of spinless electrons $\oa{\rho}{\qq}$ is obtained from Eq.\ \eqref{densop} by neglecting the pseudospin indices. This simplified U(1) model of fully valley-polarized graphene which is described by Eq.\ \eqref{u1graphene} is called \emph{U(1)-graphene} in the remainder of the paper.
Now, if one substitutes into Eq.\ \eqref{u1graphene} the non-relativistic form-factor $F_N(\qq)$, one obtains the usual single-layer quantum Hall 2DEG for spin-polarized electrons in GaAs.

The Hartree-Fock approximation applied to the graphene interaction term in Eq.\ \eqref{hint} yields\cite{cote.brey.macdonald,zhang.joglekar}
\begin{align}
\label{hhf}
\nonumber\oHhf &=\LLdeg\Vc\sum_{\idf,\QQ}\Bigl\{\left[H(\QQ)-X^{\idf\idf}(\QQ)\right]\oiia{\rho}{\idf}{\idf}{\QQ}\\
&\quad -X^{\idf\iidf}(\QQ)\oiia{\rho}{\iidf}{\idf}{\QQ}\Bigr\},
\end{align}
where $\iidf=-\idf$, and  $\QQ$'s are the reciprocal wavevectors of the WC lattice. The Hartree and Fock effective interaction potentials read, respectively,
\begin{align}
\label{hfpot}
H(\QQ)=&\frac{e^{-Q^2/2}}{Q}|\Fn{N}{\QQ}|^2\rho(-\QQ)(1-\delta_{\QQ,0}),\\
X^{\idf\idf'}(\QQ)=&\int_{0}^{\infty}dxe^{-\frac{x^2}{2}}|\Fn{N}{\QQ}|^2J_0(xQ)\rho_{\idf,\idf'}(-\QQ),
\end{align}
where $Q\equiv|\QQ|$, $J_0$ is a Bessel function, and the density averages are $\rho_{\idf,\idf'}(\QQ)=\av{\oiia{\rho}{\idf}{\idf'}{\QQ}}$,\ $\rho(\QQ)=\sum_{\idf}\rho_{\idf,\idf}(\QQ)$. We assume a triangular electron lattice for the broken-symmetry state, with reciprocal lattice vectors given by
\begin{equation}
 \QQ=Q_0\left(\frac{n}{2},\frac{n}{2}+\frac{m\sqrt{3}}{2}\right),\qquad n,m\in\mathbb{Z}.
\end{equation}
Here $Q_0$ is the length of the basis vector of the reciprocal lattice,
\begin{equation}
\label{define.Qlength}
 Q_0=\invML\left({\frac{4\pi\pnu}{\sqrt{3}\Me}}\right)^{1/2},
\end{equation}
$\pnu$ is the filling factor of the last partially filled LL,\ and $\Me$ is the number of electrons per site ($\Me=1$ corresponds to the WC,\ and $\Me\geq{2}$ to an electron-bubble crystal with $\Me$ electrons per bubble).
The single-particle Green's function in the imaginary-time Matsubara formalism \cite{mahan} reads
\begin{align}
\nonumber G_{\idf_1,\idf_2}(\QQ,\iwn)&=-\invg\int_0^{{\hbar}/{k_B T}}d\tau\exp(\iwn\tau)\\
\nonumber&\quad\times\sum_{X}\exp\left[-iQ_xX+\frac{\MLS Q_xQ_y}{2}\right]\\
&\quad\times\av{\mathcal{T}_{\tau}\oiia{c}{X-\MLS Q_y}{\idf_1}{\tau}\oiida{c}{X}{\idf_2}{0}},
\end{align}
where $T$ is the temperature, $k_B$ is the Boltzmann constant,\ $\mathcal{T}_{\tau}$ denotes imaginary-time ordering, and $\omega_n=\pi(2n+1)k_B T/\hbar$ are the Matsubara frequencies. $G_{\idf_1,\idf_2}(\QQ,\iwn)$ may be determined self-consistently from the quadratic Hamiltonian \eqref{hhf} by using the Heisenberg equations of motion within the iterative-solution method proposed in Ref. \onlinecite{MC}\ which we adopt in the present work.
\newline\indent After analytic continuation to real frequencies $\iwn\rightarrow \omega+i0^+$, $G_{\idf_1,\idf_2}(\QQ,\iwn)$ yields the retarded Green's function which may be used to calculate the DOS $g(\omega)$,
\begin{equation}
 g(\omega)=-\invg\pi\sum_{\idf}\text{Im}\ G_{\idf,\idf}(\QQ=0,\iwn\rightarrow\omega+i0^+),
\end{equation}
 and the LDOS $A(\rr,\omega)$,
\begin{equation}
 A(\rr,\omega)=-\invg\pi\sum_{\idf}\text{Im}\ G_{\idf,\idf}(\rr,\iwn\rightarrow\omega+i0^+),
\end{equation}
where the Green's function in real space reads
\begin{align}
\nonumber G_{\idf_1,\idf_2}(\rr,\iwn)&=(2\pi{\MLS})^{-1}\sum_{\QQ,\idf}\exp(-i\QQ\cdot{\rr})\Fn{N}{-\QQ}\\
&\quad\times G_{\idf_1,\idf_2}(\QQ,\iwn).
\end{align}

\section{Results and discussions}
\label{sec.results}
In this section, we discuss the spectroscopic properties of the electron-solid phases for the LL $N=2$. As already mentioned in the introduction, we concentrate on this LL for illustration purposes and because they are mostly significant from the physical point of view. We have obtained similar results for $N=1$, 3 and 4 (not discussed here).

We have chosen three different electron-solid lattices which have the same ratio $\pnu/\Me=0.14\approx{1/7}$, and hence the same lattice period given by Eq.\ \eqref{define.Qlength}.\ Our choice for the $\pnu/\Me$ ratio is rather arbitrary. We note that the $\Me=1$ and $\Me=2$ states yield in graphene the global energy minima, while the $\Me=3$ does not. It has been shown in Ref. \onlinecite{zhang.joglekar} that the ground state of graphene at $\pnu\leq{0.43}$ is an anisotropic Wigner crystal whereas at $0.28\leq\pnu\leq{0.43}$ the ground state is  the $\Me=2$ bubble crystal, and at $\pnu\leq{0.28}$ the Wigner crystal yields the lowest energy ($\Me=1$). Therefore, the $\Me=3$ phase is not the lowest-energy state in graphene; nevertheless, it is useful to analyze on the same footing all three cases $\Me=1,2,3$.

For completeness, we mention that we have also calculated the cohesive energies of other types of electron-crystal phases (not only triangular bubble phases, but also anisotropic Wigner crystals). Our results for the energies coincide with those of Ref.\ \onlinecite{zhang.joglekar} with excellent accuracy and, therefore, corroborate the DOS and LDOS results discussed below.

\subsection{(Integrated) Density of states}
\begin{figure}[htb]
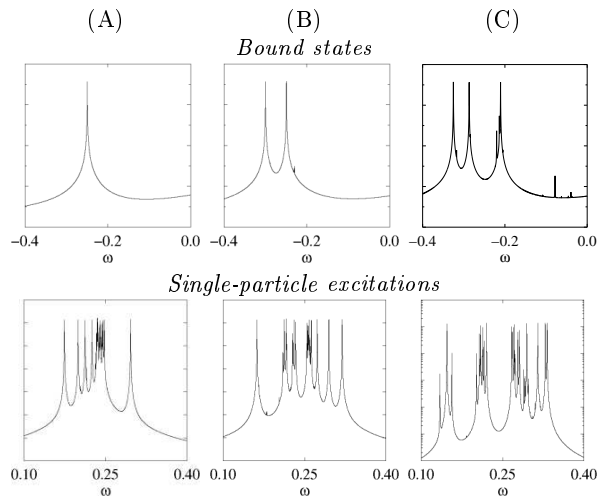

\begin{center}
\begin{tabular}{ccc}
(A) & (B) & (C)  \\
\multicolumn{3}{c}{\textit{Bound states}} \\
 \IG{\DOSsize}{pics}{DOS.LE.N.2.SU2.GR.nu.0.14.M.1.eps} &
  \IG{\DOSsize}{pics}{DOS.LE.N.2.SU2.GR.nu.0.28.M.2.eps} &
  \IG{\DOSsize}{pics}{DOS.LE.N.2.SU2.GR.nu.0.42.M.3.eps} \\ 
\multicolumn{3}{c}{\textit{Single-particle excitations}} \\
  \IG{\DOSsize}{pics}{DOS.HE.N.2.SU2.GR.nu.0.14.M.1.eps} &
  \IG{\DOSsize}{pics}{DOS.HE.N.2.SU2.GR.nu.0.28.M.2.eps} &
  \IG{\DOSsize}{pics}{DOS.HE.N.2.SU2.GR.nu.0.42.M.3.eps} \\
\end{tabular}
\end{center}
\caption{\label{DOS.Fig} Logarithmic plots for the density of states $g(\omega)$ of graphene in the $N=2$ Landau level at 
(A) $\Me=1,\ \pnu=0.14$; 
(B) $\Me=2,\ \pnu=0.28$; 
(C) $\Me=3,\ \pnu=0.42$. The frequency $\omega$ is given in units of the Coulomb scale $V_C$. The $\omega=0$ frequency is the position of the chemical potential (Fermi energy) $\mu$. The infinitesimal imaginary frequency shift $\omega\rightarrow\omega+i0^+$ is approximated by $\omega\rightarrow\omega+i\delta_{\omega}$, with $\delta_{\omega}=\pww{-4}$.\ Figures (A)-(C) in the first row yield the DOS in the low-energy frequency range (bound states of electrons), while in the second row the high-energy peaks of the DOS correspond to single-electron excitations above the ground state of the lattice.}
\end{figure}

Our results for the DOS in graphene at $N=2$ are presented in Fig.\ \ref{DOS.Fig}.\ 

We find that the DOS consists of two well-separated classes of peaks: well-defined low-energy peaks are found below the chemical potential $\mu$, which is shifted to zero energy, 
whereas the large number of peaks above $\mu$ are not that easily distinguished. 
We note here that in graphene the number of low-energy peaks in all cases is equal to $\Me$, the number of electrons in a bubble. The same result has been obtained before in Hartree-Fock studies of the simpler single-layer 2D quantum Hall system.\cite{cote.bubbles}\ We checked that the same property holds true also for U(1)-graphene, and in non-relativistic two-component quantum Hall systems, such as a bilayer with zero layer separation.

In the simpler single-layer 2D quantum Hall system in GaAs, the DOS at $\Me=1$ exhibits the features of the Hofstadter butterfly structure.\cite{cote.bubbles} It means the following: given that the filling factor may be represented by a ratio of two integers $p,\ q$ without a common divisor, $\pnu=p/q$, these integers $p,\ q$ determine then the structure of the single-particle energy spectrum of the system; namely, there should exist $p$ low-energy levels and $q-p$ high-energy levels (Hofstadter butterfly counting rule). In the DOS, which is a function of frequency, these energy levels are recognized as smoothed peaks.
The Hofstadter butterfly counting rule was confirmed for $\Me=1$ in the single-layer 2DEG in GaAs, while for $\Me\geq{2}$ it is claimed that counting of the single-particle levels is different: the number of low-energy peaks is equal to $\Me$, whereas nothing is known about what is the precise rule for counting the number of high-energy peaks.\cite{cote.bubbles}

What will be important in the following discussion is the \emph{order of indexing of the DOS peaks}. We will count the peaks in the DOS with respect to \emph{increasing the frequency}\ $\omega$. In Fig.\ \ref{DOS.Fig}(A),\ the \emph{first} DOS peak is obviously the lowest-energy one with energy $\approx-0.24\Vc$. The \emph{second peak} in (A) is a higher-energy one with energy $\approx0.17\Vc$. The numbering of peaks continues until we reach the utmost-right peak with energy $\approx0.3\Vc$. The same indexing rule is applied to the cases (B) and (C).
One should note that in all three cases the DOS peaks with the same index may have rather different energies: while in (A) the second DOS
peak belongs already to the high-energy region, in (B) it still lies below the Fermi level.

In addition, one should have a procedure of extracting the energies of the DOS peaks from the smoothed DOS vs frequency dependence shown in Fig.\ \ref{DOS.Fig}. In the low-energy regime it may be always done reliably. In the high-energy regime, however, there is a larger number of closely-located DOS peaks, the shapes, widths, and amplitudes of which depend
sensitively on the imaginary frequency shift $\delta_\omega$. The latter is used for the analytical continuation into the upper complex half-plane of the Green's function, $i\omega_n\rightarrow\omega+i\delta_\omega$. Physically, this imaginary frequency shift represents a level broadening due, e.g., to disorder. 
We have found that the best way to extract only those peaks which are physical is to place a cutoff $\Delta$ on the DOS peak amplitude, so that peaks with amplitude less than $g_{\rm{max}}\Delta$ are neglected, with $g_{\rm{max}}\propto \delta_{\omega}^{-1}$ the maximum peak amplitude. In our study, we have chosen $\delta_\omega=\pww{-4}$,\ and\  $\Delta=0.5$. The number of shells of reciprocal lattice vectors $\QQ$'s is $\nsh=8$, so that the actual number of vectors is $\nQ=241$. Single-particle energies which are extracted from the smoothed DOS, will be used below in the calculation of the LDOS.
In the U(1)-graphene and the single-layer cases at the same densities considered here, we are able to extract $7$ DOS peaks; at $\pnu=0.14,\ \Me=1$, there is one lowest-energy peak, and $6$ high-energy ones.  These are exactly the numbers of single-particle levels dictated by the Hofstadter butterfly counting rule.\cite{cote.bubbles}

 In graphene and the quantum Hall bilayer, studied within the two-component model, we obtain the number of identified DOS peaks around $14$ [with deviation of not more than one wrongly identified peak].\ Due to the additional SU(2)-symmetry in the latter two cases, it is natural to expect that the number of single-particle levels is thus \emph{doubled}.
\subsection{Real-space density profile}
\begin{figure}[htb]
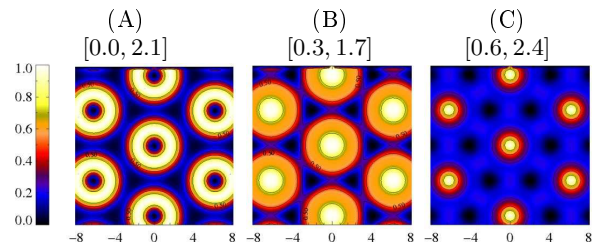

\begin{center}
 \begin{tabular}{ccc}
  (A) & (B) & (C) \\
$[0.0,2.1]$ & $[0.3,1.7]$ & $[0.6,2.4]$ \\
  \IG{\NRBARsize}{pics}{BAR.NR.N.2.SU2.GR.nu.0.14.M.1.eps} &
  \IG{\NRsize}{pics}{NR.N.2.SU2.GR.nu.0.28.M.2.eps} &
  \IG{\NRsize}{pics}{NR.N.2.SU2.GR.nu.0.42.M.3.eps} \\
 \end{tabular}
\end{center}
\caption{\label{NR.Fig} (Color online) Real-space density profile $n(\rr)$ in graphene in the $N=2$ LL. Choices (A)-(C) are the same as in Fig.\ \ref{DOS.Fig}. Minima and maxima of $n(\rr)$ written inside the square brackets as [min,\ max] correspond to the values [0.0,\ 1.0] in the colour plots (blue and white colours, correspondingly).}
\end{figure}

For later comparison with our results for the LDOS, we calculated the real-space electron density profile
\begin{equation}
\label{def.NR}
n(\rr)=\frac{1}{2\pi{\MLS}}\sum_{\QQ}\exp(-i\QQ\cdot{\rr})\Fn{N}{-\QQ}\rho(\QQ)
\end{equation} 
for the same choices of ($\pnu,\ \Me$)\ as in Fig.\ \ref{DOS.Fig}. The results, which are shown in Fig.\ \ref{NR.Fig}, agree with previous calculations for graphene performed by Zhang and Joglekar.\cite{zhang.joglekar}
\subsection{Local density of states}
\label{subsec.LDOS}
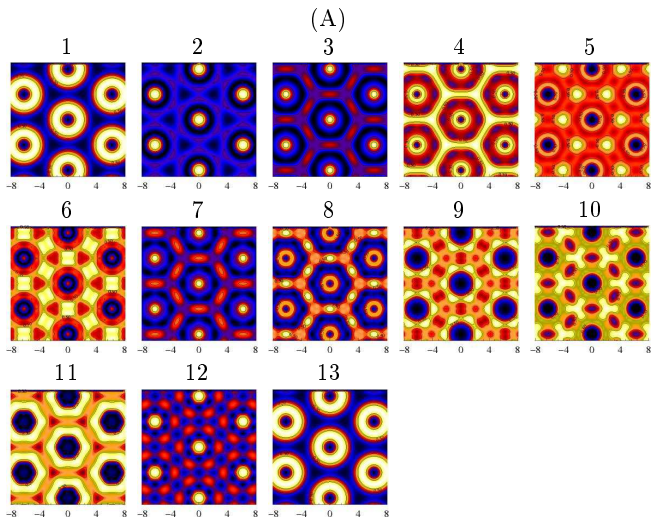
\begin{figure}[htb]
\begin{center}
\input{FIG.LDOS.SU2.GR.nu.0.14.M.1.tex}
\end{center}
\caption{\label{LDOS.Fig.A}(Color online) LDOS $A(\rr,\omega)$ for graphene at $\pnu=0.14,\ \Me=1$ [case (A)]. Contour colours are graded in the same way as defined in Fig.\ \ref{NR.Fig}. The contour plots are ordered with respect to the index of DOS peaks [indicated above the plots]. The number of extracted DOS peaks is $\Np$=13.} 
\end{figure}

Our results for the LDOS in graphene are presented in Figs.\ \ref{LDOS.Fig.A},\ \ref{LDOS.Fig.B},\ and \ref{LDOS.Fig.C}.\ The LDOS patterns is plotted for all three cases (A)-(C), as in Fig.\ \ref{DOS.Fig}, and at the energies of all extracted single-particle DOS peaks situated in increasing order.\ We obtain that the \emph{rescaled} (to the range of $[0.0,\ 1.0]$) real-space patterns of $A(\rr,\omega)$, calculated at the \emph{first} \emph{four} DOS peaks for all three choices of $(\pnu,\ \Me)$, coincide among themselves. There is also an approximate mapping between the LDOS patterns at the fifth DOS peak, although less pronounced than for the first four ones (one sees correspondence between the positions of maxima and minima, but the colors deviate slightly in each case). For larger values of the peak index, we start to see considerable discrepancies between the LDOS patterns. Also the number of extracted peaks $\Np$ is different for each case. The latter property is due to the very approximate nature of our extraction procedure: while low-energy peaks are always identified reliably, the high-energy peaks are determined only approximately. However, the accuracy is quite good.
We also note a very interesting property of the LDOS at the last two peaks for (B)-(C): the LDOS patterns are identical, but their positions are \emph{swapped}. We do not have any physical argument why this should be the case, but it could be a feature that appears when $M > N$. However, this statement is a mere speculation and a more detailed investigation is required to clarify this aspect. 

\begin{figure}[htb]
\begin{center}
\input{FIG.LDOS.SU2.GR.nu.0.28.M.2.tex}
\end{center}
\caption{\label{LDOS.Fig.B}(Color online) LDOS $A(\rr,\omega)$ for graphene at $\pnu=0.28,\ \Me=2$ [case (B)]. $\Np=15$.} 
\end{figure}
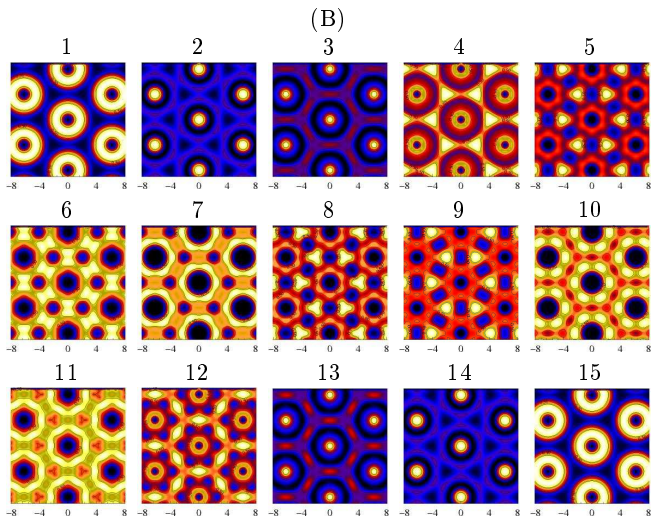
\begin{figure}[htb]
\begin{center}
\input{FIG.LDOS.SU2.GR.nu.0.42.M.3.tex}
\end{center}
\caption{\label{LDOS.Fig.C}(Color online) LDOS $A(\rr,\omega)$ for graphene at $\pnu=0.42,\ \Me=3$ [case (C)]. $\Np$=14.} 
\end{figure}
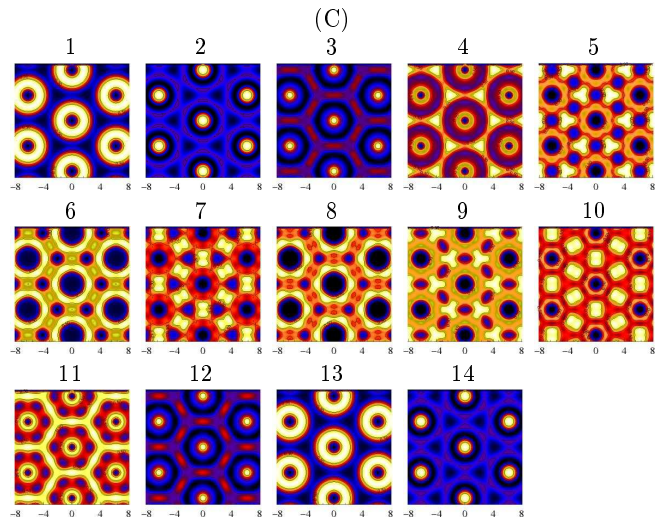

\subsection{Comparison with the real-space density}
Now, to compare the LDOS patterns shown in Figs.\ \ref{LDOS.Fig.A},\ \ref{LDOS.Fig.B},\ \ref{LDOS.Fig.C} with the real-space density profile $n(\rr)$ defined in Eq.\ \eqref{def.NR} and plotted in Fig.\ \ref{NR.Fig}, we introduce the \emph{resummed} LDOS $\tilde{A}(\rr,\omega)$, defined for a fixed single-particle energy $\omega_i$ as a sum of all LDOS patterns at smaller peak energies,
\begin{equation}
\tilde{A}(\rr,\omega_i)=\sum_{j=1}^{i}A(\rr,\omega_j).
\end{equation}

\begin{figure}[htb]
 \begin{center}
\input{FIG.RLDOS.SU2.GR.tex}
 \end{center}
 \caption{\label{RLDOS.Fig}(Color online) Resummed LDOS $\tilde{A}(\rr,\omega)$ for graphene at the three first DOS peaks for (A) [$\pnu=0.14,\ \Me=1$]. Contour colours are graded in the same way as defined in Fig.\ \ref{NR.Fig}.}
\end{figure}
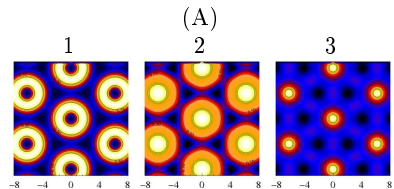
Given the excellent coincidence of the LDOS patterns shown in Fig.\ \ref{RLDOS.Fig} with the real-space densities in Fig.\ \ref{NR.Fig},\ one may empirically write 
\begin{equation}
 n(\rr,\Me=i)\leftrightarrow\tilde{A}(\rr,\omega_i),\ i=1,2,3,
\end{equation}
where the sign $\leftrightarrow$ means mapping between the \emph{rescaled to the [0.0,1.0] interval} quantities.
This means that the real-space density of the $\Me$-electron bubble crystal is determined by the sum of the LDOS at the 
$\Me$ negative-energy peaks. More surprisingly,
because of the correspondence between the LDOS patterns of the low-energy peaks for all different $\Me$ bubble crystals, one
may determine the real-space density pattern of the $\Me$ bubble crystal by summing the LDOS of the $\Me$ peaks of lowest energy
for {\sl any} of the electron-solid phases -- the LDOS patterns of the $\Me=1$ Wigner crystal, e.g., contains thus the information
of the density of all other $\Me$ bubble crystals.

\subsection{U(1)-graphene: local density of states}
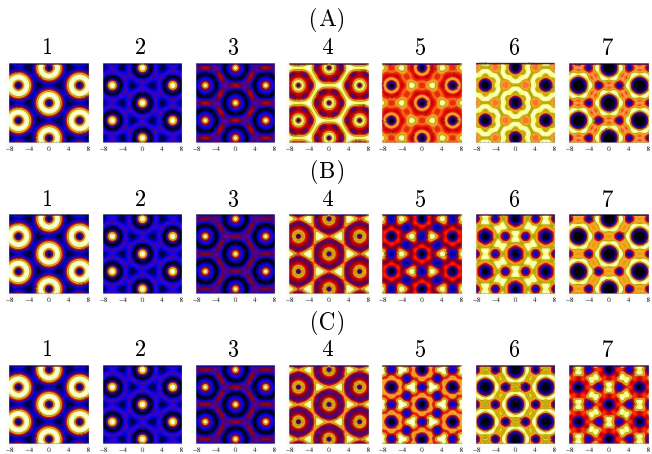
\begin{figure}[ht]
\begin{center}
 \input{FIG.LDOS.U1.GR.tex}
\end{center}
\caption{\label{LDOS.U(1).Fig}(Color online) LDOS $A(\rr,\omega)$ for U(1)-graphene. The number of extracted DOS peaks $\Np=7$.}
\end{figure}

In Fig.\ \ref{LDOS.U(1).Fig} we present the LDOS for U(1)-graphene. The number of extracted DOS peaks for all three density choices (A)-(C) is $\Np=7$. This is in accordance with the Hofstadter butterfly counting rule. We also see an excellent correspondence of the LDOS for the first four DOS peaks, then, also a good coincidence of the LDOS at the peaks 6-7 for (A)-(B), whereas these two LDOS patterns for (C) interchange places,
as compared with the patterns 6-7 for (A)-(B). This interchange phenomenon is the same as observed for graphene and is not yet understood. 

In general, the U(1)-graphene results coincide numerically with those for graphene when one takes into account the SU(2) symmetry for the valley pseudospin. This indicates that one may use the U(1)-model instead of the more complex SU(2)-symmetric one for the discussion of the density patterns, hence simplifying further calculations on graphene. Moreover, it indicates that in the electron-crystal phases considered above the valley degree of freedom is fully polarized.

\section{Conclusions}
\label{sec.concl.}
The aim of this work is to show how a high-field electron-solid phase in the 2DEG may be detected by optical means in graphene.
We have calculated the DOS and the LDOS of electron-solid phases in the Hartree-Fock approximation in the $N=2$ LL.
\newline\indent We show that the number of low-energy DOS peaks in graphene is given by the number of electrons per site $\Me$. 
This result is similar to the previous DOS calculation in the Hartree-Fock approximation for GaAs.\cite{cote.bubbles}
\newline\indent We found that the rescaled LDOS is identical for different filling factors $\pnu$, as long as the ratio $\pnu/\Me$, which determines the lattice spacing of the $\Me$-electron bubble crystal, is kept fixed, and the LDOS frequency is taken at the DOS peak with the same index (for the first four indices). In particular, this result yields an unexpected conclusion that, e.g. by fixing the filling factor $\pnu=0.14$ in the $N=2$ LL,\ and using STM, one could observe in the LDOS the whole succession of electron-crystal density patterns with $\Me=1,\ 2,\ 3$ by fixing the applied STM voltage at the consecutive first three single-particle excitation energies, and summing up the LDOS to obtain the resummed LDOS $\tilde{A}(\rr,\omega)$.
\newline\indent We believe that this LDOS correspondence holds true for all single-particle excitations resolved as individual DOS peaks so far (accounting for interchanging of the last two peaks in the $\Me=2,3$ cases) and for all LL's (similar conclusions follow from our calculations in the LLs $N=1,\ 3,\ 4$).

We also obtained the same LDOS correspondence for other models of the 2DEG: (i) in a single-layer GaAs heterostructure; (ii) U(1)-graphene; (iii) bilayer. This implies that the observed LDOS $\pnu/\Me$ scaling is independent of the underlying interaction potential and the number of inner discrete degrees of freedom. The fact that the U(1)-graphene results coincide numerically with those for graphene indicates that the electron crystals considered here are completely valley-pseudospin polarized.

\begin{acknowledgements}
\noindent We thank P.\ Lederer for fruitful discussions.
O.\ P.\ acknowledges financial support from the European Commission through the Marie-Curie Foundation contract MEST CT 2004-51-4307, 
and from the Gates Cambridge Scholarship Trust. The work of C.M.S. was partially supported by the Netherlands Organization for Scientific Research (NWO). 
\end{acknowledgements}

\end{document}

%% file: newcommands.tex
\newcommand{\pww}[1]{{10^{#1}}}

\newcommand{\ML}{l_{B}}
\newcommand{\invML}{l_{B}^{-1}}
\newcommand{\MLS}{{l_{B}^2}}

\newcommand{\affCU}{Faculty of Mathematics, Wilberforce Road, University of Cambridge, CB3 0WA, United Kingdom}
\newcommand{\affLPS}{Laboratoire de Physique des Solides, CNRS UMR 8502, Universit\'{e} Paris Sud, F-91405 Orsay Cedex, France}
\newcommand{\affUU}{Institute for Theoretical Physics, University of
Utrecht, Leuvenlaan 4, 3584 CE Utrecht, The Netherlands}

\newcommand{\IG}[3]{\includegraphics[width=#1]{#2/#3}}

\newcommand{\btwocols}{\begin{multicols}{2}}
\newcommand{\etwocols}{\end{multicols}}

\newcommand{\Np}{N_{\rm{p}}}

\newcommand{\journ}[4]{#1\textbf{#2},\ #3\ (#4)}

\newcommand{\Fn}[2]{\mathcal{F}_{#1}(#2)}

\newcommand{\invg}{{N_\phi^{-1}}}

\newcommand{\oHint}{\hat{\mathcal{H}}_{\rm{int}}}
\newcommand{\oHhf}{\hat{\mathcal{H}}_{\rm{int}}^{\rm{(HF)}}}

\newcommand{\av}[1]{\langle{#1}\rangle}

\newcommand{\oii}[3]{\hat{#1}_{#2,#3}}

\newcommand{\oa}[2]{\hat{#1}(#2)}

\newcommand{\oiia}[4]{\hat{#1}_{#2,#3}(#4)}

\newcommand{\oiid}[3]{\hat{#1}_{#2,#3}^{\dagger}}

\newcommand{\oiida}[4]{\hat{#1}_{#2,#3}^{\dagger}(#4)}

\newcommand{\qq}{\textbf{q}}

\newcommand{\LLdeg}{{N_{\rm{\phi}}}}
\newcommand{\Vc}{V_{\rm{C}}}

\newcommand{\nsh}{N_{\rm{sh}}}

\newcommand{\pnu}{\nu_{N}}

\newcommand{\nQ}{N_{Q}}

\newcommand{\Me}{M_{\rm{e}}}

\newcommand{\idf}{{\beta}}
\newcommand{\iidf}{\bar{\idf}}

\newcommand{\be}{\begin{equation}}
\newcommand{\ee}{\end{equation}}
\newcommand{\bd}{\begin{displaymath}}
\newcommand{\ed}{\end{displaymath}}

\newcommand{\iwn}{i\omega_n}

 \newcommand{\QQ}{\textbf{Q}}

\newcommand{\rr}{\textbf{r}}

\newcommand{\authline}{}
 \newcommand{\authcote}{{R.\ C\^{o}t\'{e}}}

\newcommand{\jpr}{\ Phys.\ Rev.\ }
 \newcommand{\jprb}{{\ Phys.\ Rev.\ B\ }}
 \newcommand{\jprl}{{\ Phys.\ Rev.\ Lett.\ }}
 
 \newcommand{\jrmp}{{\ Rev.\ Mod.\ Phys.\ }}

%% file: FIG.LDOS.SU2.GR.nu.0.14.M.1.tex
\begin{tabular}{*{5}{c}}
\multicolumn{5}{c}{(A)} \\
1 & 2 & 3 & 4 & 5 \\
 \IG{\LDOSsize}{pics}{NOBAR.LDOS.N.2.SU2.GR.nu.0.14.M.1.P1.eps} & 
 \IG{\LDOSsize}{pics}{LDOS.N.2.SU2.GR.nu.0.14.M.1.P2.eps} &
 \IG{\LDOSsize}{pics}{LDOS.N.2.SU2.GR.nu.0.14.M.1.P3.eps} & 
 \IG{\LDOSsize}{pics}{LDOS.N.2.SU2.GR.nu.0.14.M.1.P4.eps} &
 \IG{\LDOSsize}{pics}{LDOS.N.2.SU2.GR.nu.0.14.M.1.P5.eps} \\
6 & 7 & 8 & 9 & 10 \\
 \IG{\LDOSsize}{pics}{LDOS.N.2.SU2.GR.nu.0.14.M.1.P6.eps} &
 \IG{\LDOSsize}{pics}{LDOS.N.2.SU2.GR.nu.0.14.M.1.P7.eps} &
 \IG{\LDOSsize}{pics}{LDOS.N.2.SU2.GR.nu.0.14.M.1.P8.eps} &
 \IG{\LDOSsize}{pics}{LDOS.N.2.SU2.GR.nu.0.14.M.1.P9.eps} &
 \IG{\LDOSsize}{pics}{LDOS.N.2.SU2.GR.nu.0.14.M.1.P10.eps} \\
11 & 12 & 13 \\
 \IG{\LDOSsize}{pics}{LDOS.N.2.SU2.GR.nu.0.14.M.1.P11.eps} &
 \IG{\LDOSsize}{pics}{LDOS.N.2.SU2.GR.nu.0.14.M.1.P12.eps} &
 \IG{\LDOSsize}{pics}{LDOS.N.2.SU2.GR.nu.0.14.M.1.P13.eps}  
\end{tabular}

%% file: FIG.LDOS.SU2.GR.nu.0.28.M.2.tex
\begin{tabular}{*{5}{c}}
 \multicolumn{5}{c}{(B)} \\
1 & 2 & 3 & 4 & 5 \\
 \IG{\LDOSsize}{pics}{NOBAR.LDOS.N.2.SU2.GR.nu.0.28.M.2.P1.eps} & 
 \IG{\LDOSsize}{pics}{LDOS.N.2.SU2.GR.nu.0.28.M.2.P2.eps} &
 \IG{\LDOSsize}{pics}{LDOS.N.2.SU2.GR.nu.0.28.M.2.P3.eps} & 
 \IG{\LDOSsize}{pics}{LDOS.N.2.SU2.GR.nu.0.28.M.2.P4.eps} &
 \IG{\LDOSsize}{pics}{LDOS.N.2.SU2.GR.nu.0.28.M.2.P5.eps} \\
6 & 7 & 8 & 9 & 10 \\
 \IG{\LDOSsize}{pics}{LDOS.N.2.SU2.GR.nu.0.28.M.2.P6.eps} &
 \IG{\LDOSsize}{pics}{LDOS.N.2.SU2.GR.nu.0.28.M.2.P7.eps} &
 \IG{\LDOSsize}{pics}{LDOS.N.2.SU2.GR.nu.0.28.M.2.P8.eps} & 
 \IG{\LDOSsize}{pics}{LDOS.N.2.SU2.GR.nu.0.28.M.2.P9.eps} & 
 \IG{\LDOSsize}{pics}{LDOS.N.2.SU2.GR.nu.0.28.M.2.P10.eps} \\
11 & 12 & 13 & 14 & 15 \\
 \IG{\LDOSsize}{pics}{LDOS.N.2.SU2.GR.nu.0.28.M.2.P11.eps} & 
 \IG{\LDOSsize}{pics}{LDOS.N.2.SU2.GR.nu.0.28.M.2.P12.eps} & 
 \IG{\LDOSsize}{pics}{LDOS.N.2.SU2.GR.nu.0.28.M.2.P13.eps} &
 \IG{\LDOSsize}{pics}{LDOS.N.2.SU2.GR.nu.0.28.M.2.P14.eps} &
 \IG{\LDOSsize}{pics}{LDOS.N.2.SU2.GR.nu.0.28.M.2.P15.eps} 
\end{tabular}

%% file: FIG.LDOS.SU2.GR.nu.0.42.M.3.tex
\begin{tabular}{*{5}{c}}
 \multicolumn{5}{c}{(C)} \\
1 & 2 & 3 & 4 & 5 \\
 \IG{\LDOSsize}{pics}{NOBAR.LDOS.N.2.SU2.GR.nu.0.42.M.3.P1.eps} & 
 \IG{\LDOSsize}{pics}{LDOS.N.2.SU2.GR.nu.0.42.M.3.P2.eps} &
 \IG{\LDOSsize}{pics}{LDOS.N.2.SU2.GR.nu.0.42.M.3.P3.eps} & 
 \IG{\LDOSsize}{pics}{LDOS.N.2.SU2.GR.nu.0.42.M.3.P4.eps} &
 \IG{\LDOSsize}{pics}{LDOS.N.2.SU2.GR.nu.0.42.M.3.P5.eps} \\
6 & 7 & 8 & 9 & 10 \\
 \IG{\LDOSsize}{pics}{LDOS.N.2.SU2.GR.nu.0.42.M.3.P6.eps} &
 \IG{\LDOSsize}{pics}{LDOS.N.2.SU2.GR.nu.0.42.M.3.P7.eps} &
 \IG{\LDOSsize}{pics}{LDOS.N.2.SU2.GR.nu.0.42.M.3.P8.eps} &
 \IG{\LDOSsize}{pics}{LDOS.N.2.SU2.GR.nu.0.42.M.3.P9.eps} & 
 \IG{\LDOSsize}{pics}{LDOS.N.2.SU2.GR.nu.0.42.M.3.P10.eps} \\
11 & 12 & 13 & 14 \\
 \IG{\LDOSsize}{pics}{LDOS.N.2.SU2.GR.nu.0.42.M.3.P11.eps} & 
 \IG{\LDOSsize}{pics}{LDOS.N.2.SU2.GR.nu.0.42.M.3.P12.eps} &
 \IG{\LDOSsize}{pics}{LDOS.N.2.SU2.GR.nu.0.42.M.3.P13.eps} &
 \IG{\LDOSsize}{pics}{LDOS.N.2.SU2.GR.nu.0.42.M.3.P14.eps} 
\end{tabular}

%% file: FIG.RLDOS.SU2.GR.tex
\begin{tabular}{*{3}{c}}
\multicolumn{3}{c}{(A)} \\
1 & 2 & 3 \\
 \IG{\LDOSsize}{pics}{RLDOS.N.2.SU2.GR.nu.0.14.M.1.P1.eps} & 
 \IG{\LDOSsize}{pics}{RLDOS.N.2.SU2.GR.nu.0.14.M.1.P2.eps} &
 \IG{\LDOSsize}{pics}{RLDOS.N.2.SU2.GR.nu.0.14.M.1.P3.eps} 
\end{tabular}

%% file: FIG.LDOS.U1.GR.tex
\begin{tabular}{*{7}{c}}
\multicolumn{7}{c}{(A)} \\
1 & 2 & 3 & 4 & 5 & 6 & 7 \\
 \IG{\LDOSsizeU}{pics}{LDOS.N.2.U1.GR.nu.0.14.M.1.P1.eps} & 
 \IG{\LDOSsizeU}{pics}{LDOS.N.2.U1.GR.nu.0.14.M.1.P2.eps} &
 \IG{\LDOSsizeU}{pics}{LDOS.N.2.U1.GR.nu.0.14.M.1.P3.eps} & 
 \IG{\LDOSsizeU}{pics}{LDOS.N.2.U1.GR.nu.0.14.M.1.P4.eps} &
 \IG{\LDOSsizeU}{pics}{LDOS.N.2.U1.GR.nu.0.14.M.1.P5.eps} &
 \IG{\LDOSsizeU}{pics}{LDOS.N.2.U1.GR.nu.0.14.M.1.P6.eps} &
 \IG{\LDOSsizeU}{pics}{LDOS.N.2.U1.GR.nu.0.14.M.1.P7.eps} \\
\multicolumn{7}{c}{(B)} \\
1 & 2 & 3 & 4 & 5 & 6 & 7 \\
 \IG{\LDOSsizeU}{pics}{LDOS.N.2.U1.GR.nu.0.28.M.2.P1.eps} &
 \IG{\LDOSsizeU}{pics}{LDOS.N.2.U1.GR.nu.0.28.M.2.P2.eps} &
 \IG{\LDOSsizeU}{pics}{LDOS.N.2.U1.GR.nu.0.28.M.2.P3.eps} &
 \IG{\LDOSsizeU}{pics}{LDOS.N.2.U1.GR.nu.0.28.M.2.P4.eps} &
 \IG{\LDOSsizeU}{pics}{LDOS.N.2.U1.GR.nu.0.28.M.2.P5.eps} &
 \IG{\LDOSsizeU}{pics}{LDOS.N.2.U1.GR.nu.0.28.M.2.P6.eps} & 
 \IG{\LDOSsizeU}{pics}{LDOS.N.2.U1.GR.nu.0.28.M.2.P7.eps} \\
\multicolumn{7}{c}{(C)} \\
1 & 2 & 3 & 4 & 5 & 6 & 7 \\
 \IG{\LDOSsizeU}{pics}{LDOS.N.2.U1.GR.nu.0.42.M.3.P1.eps} &
 \IG{\LDOSsizeU}{pics}{LDOS.N.2.U1.GR.nu.0.42.M.3.P2.eps} &
 \IG{\LDOSsizeU}{pics}{LDOS.N.2.U1.GR.nu.0.42.M.3.P3.eps} &
 \IG{\LDOSsizeU}{pics}{LDOS.N.2.U1.GR.nu.0.42.M.3.P4.eps} &
 \IG{\LDOSsizeU}{pics}{LDOS.N.2.U1.GR.nu.0.42.M.3.P5.eps} &
 \IG{\LDOSsizeU}{pics}{LDOS.N.2.U1.GR.nu.0.42.M.3.P6.eps} & 
 \IG{\LDOSsizeU}{pics}{LDOS.N.2.U1.GR.nu.0.42.M.3.P7.eps} \\
\end{tabular}

%% file: main.bbl
\begin{thebibliography}{00}
\bibitem{wigner} \authline{E.\ P.\ Wigner},\ \journ{\jpr}{1002}{46}{1934}.
\bibitem{FPA}H. Fukuyama, P. M. Platzman, and P. W. Anderson, \journ{\jprb}{19}{5211}{1979}.
\bibitem{yoshioka.fukuyama}D.\ Yoshioka and H.\ Fukuyama, J.\ Phys.\ Soc.\ Japan\ \textbf{47}, 394 (1979).
\bibitem{yoshioka.lee}D.\ Yoshioka and P.\ A.\ Lee,\ \journ{\jprb}{27}{4986}{1983}.
\bibitem{FL} H. Fukuyama and P. A. Lee, \journ{\jprb}{17}{535}{1977}; {\bf 18}, 6245 (1978). 
\bibitem{fogler} A.\ A.\ Koulakov,\ M.\ Fogler,\ and B.\ I.\ Shklovskii, \journ{\jprl}{76}{499}{1996};\ M.\ M.\ Fogler, A.\ A.\ Koulakov,\ and B.\ I.\ Shklovskii, \journ{\jprb}{54}{1853}{1996};\ R.\ Moessner and J.\ T.\ Chalker, \journ{\jprb}{54}{5006}{1996}.
\bibitem{morais} M.\ O.\ Goerbig,\ P.\ Lederer,\ and C.\ Morais Smith,\ \journ{\jprb}{69}{115327}{2004}.
\bibitem{cote.bubbles}\authcote, C. B. Doiron, J. Bourassa, and H. A. Fertig,\ \journ{\jprb}{68}{155327}{2004}.
\bibitem{fogler.QHLC} E. Fradkin and S. A. Kivelson, \journ{\jprb}{59}{8065}{1999}; 
M.\ M.\ Fogler, in C. Berthier, L. P. L\'evy, and G. Martinez (eds.), {\sl High Magnetic Fields}, Springer Verlag (2001).
\bibitem{FQHErev} H. L. Stormer, D. C. Tsui, and A. C. Gossard, Rev. Mod. Phys. {\bf 71}, S298 (1999).
\bibitem{eis} J.\ P.\ Eisenstein, K.\ B.\ Cooper, L.\ N.\ Pfeiffer, and K.\ W.\ West, Phys.\ Rev.\ Lett.\ {\bf 88}, 076801 (2002).
\bibitem{cooper}K. B. Cooper, M. P. Lilly, and J. P. Eisenstein, L. N. Pfeiffer and K. W. West,\journ{\jprb}{60}{R11285}{1999}. 
\bibitem{GLM} M.\ O.\ Goerbig,\ P.\ Lederer,\ and C.\ Morais Smith,\ \journ{\jprb}{68}{241302(R)}{2003}. 
\bibitem{fertig.WC} H.\ A.\  Fertig, "Properties of the Electron Solid", in "Perspectives in Quantum Hall Effects", eds. S. Das Sarma and A. Pinczuk, (Wiley, New York, 1997).
\bibitem{lewis1} R.\ M.\ Lewis, P.\ D.\ Ye, L.\ W.\ Engel, D.\ C.\ Tsui, L.\ N.\ Pfeiffer, and K.\ W.\ West, Phys.\ Rev.\ Lett.\ {\bf 89}, 136804 (2002).
\bibitem{lewis2}R.\ M.\ Lewis, Y.\ P.\ Chen, L.\ W.\ Engel, D.\ C.\ Tsui, P.\ D.\ Ye, and L.\ N.\ Pfeiffer, Phys.\ Rev.\ Lett.\ {\bf 93}, 176808 (2004).
\bibitem{lewis3}R.\ M.\ Lewis, Y.\ P.\ Chen, L.\ W.\ Engel, D.\ C.\ Tsui, L.\ N.\ Pfeiffer, and K.\ W.\ West, Phys.\ Rev.\ B\ {\bf 71}, 081301 (2005).
\bibitem{chen.LLLWC} Y. P. Chen, R. M. Lewis, L. W. Engel, D. C. Tsui, P. D. Ye, Z. H. Wang, L. N. Pfeiffer, and K. W. West,\ \journ{\jprl}{93}{206805}{2004}.
\bibitem{chen.melting} Y. P. Chen, G. Sambandamurthy, Z. H. Wang, R. M. Lewis, L. W. Engel, D. C. Tsui, P. D. Ye, L. N. Pfeiffer and K. W. West, \journ{Nature Physics\ }{2}{452}{2006}.
\bibitem{fischer.STM}{\O}. Fischer, M. Kugler, I. Maggio-Aprile, and C. Berthod,\ \journ{\jrmp}{79}{353}{2007}.
\bibitem{geim.novoselov}A.\ K.\ Geim and K.\ S.\ Novoselov,\ \journ{Nature Materials\ }{6}{183}{2007}.
\bibitem{antonioRev}For a review, see A. H. Castro Neto, N. M. R. Peres, K. S. Novoselov, and A. K. Geim, Rev. Mod. Phys. {\bf 81}, 109 (2009).
\bibitem{novoselov} K. S. Novoselov, A. K. Geim, S. V. Morozov, D. Jiang, M. I. Katsnelson, I. V. Grigorieva, S. V. Dubonos and A. A. Firsov,\ Nature (London)\ \textbf{438},\ 197\ (2005).
\bibitem{zhang} Y. Zhang, Y.-W. Tan, H. L. Stormer, and P. Kim, Nature {\bf 438} 201, (2005).
\bibitem{roomT} K. S. Novoselov, Z. Jiang, Y. Zhang, S. V. Morozov, H. L. Stormer, U. Zeitler, J. C. Maan, G. S. Boebinger, P. Kim, A. K. Geim, Science {\bf 315}, 1379 (2007). 
\bibitem{goerbig.interactions.graphene}M. O. Goerbig, R. Moessner, and B. Dou\c{c}ot, \jprb\ \textbf{74}, 161407(R) (2006).
\bibitem{AC} V. M. Apalkov and T. Chakraborty, Phys. Rev. Lett. {\bf 97}, 126801 (2006).
\bibitem{toke} C. T\"oke, P. E. Lammert, V. H. Crespi, and J. K. Jain, Phys. Rev. B {\bf 74}, 235417 (2006); C. T\"oke and J. K. Jain, Phys. Rev. B {\bf 75}, 245440 (2007).
\bibitem{khves}D. V. Khveshchenko, \journ{\jprb}{75}{153405}{2007}.
\bibitem{GR} M. O. Goerbig and N. Regnault, Phys. Rev. B {\bf 75}, 241405 (2007); R. de Gail, N. Regnault, and M. O. Goerbig, Phys. Rev. B {\bf 77}, 165310 (2008). 
\bibitem{zhang.joglekar}C.-H.\ Zhang and Yogesh\ N.\ Joglekar, \jprb\ \textbf{75},\ 245414\ (2007).
\bibitem{zhang.joglekar.LLM}C.-H.\ Zhang and Yogesh\ N.\ Joglekar,\ \jprb\ \textbf{77}, 205426\ (2008).
\bibitem{jianhui.wang}J. Wang, A. Iyengar, H. A. Fertig, and L. Brey,\ \jprb\ \textbf{78}, 165416 (2008). 
\bibitem{hao.wang}H. Wang, D. N. Sheng, L. Sheng, and F. D. M. Haldane,\ \jprl\ \textbf{100},\ 116802\ (2008).
\bibitem{cote.skyrme.wc.graphene}\authcote, J.-F. Jobidon, and H. A. Fertig,\ \jprb\ \textbf{78},\ 085309 (2008).
\bibitem{martin} J. Martin, N. Akerman, G. Ulbricht, T. Lohmann, J. H. Smet, K. von Klitzing, and A. Yacoby,
Nature Physics {\bf 4}, 144 (2008). 
\bibitem{mallet} P. Mallet, F. Varchon, C. Naud, L. Magaud, C. Berger, and J.-Y. Veuillen, Phys. Rev. B {\bf 76}, 041403 (2007).
\bibitem{eva} G. Li, A. Luican, E. Y. Andrei, arXiv:0803.4016.
\bibitem{no.wc.graphene}H.\ P.\ Dahal,\ Y.\ N.\ Joglekar,\ K.\ S.\ Bedell,\ and\ A.\ V.\ Balatsky,\ \jprb\ \textbf{74},\ 233405 (2006).
\bibitem{cote.macdonald}\authcote\ and\ A.\ H.\ MacDonald,\ \jprl\ \textbf{65}, 2662 (1990);\ \jprb\ \textbf{44}, 8759 (1991).
\bibitem{cote.brey.macdonald}\authcote,\ L.\ Brey,\ and A.\ H.\ MacDonald,\ \jprl\ \textbf{46}, 10239 (1992).
\bibitem{nomura.macdonald}K.\ Nomura and A.\ H.\ MacDonald,\ \jprl\ \textbf{96},\ 256602\ (2006).
\bibitem{AF} J. Alicea and M. P. A. Fisher, Phys. Rev. B {\bf 74}, 075422 (2006).
\bibitem{mahan} G.\ D.\ Mahan, Many-particle physics, Indiana Univ. (1980).
\bibitem{MC} \authcote\ and A.\ H.\ MacDonald, \journ{\jprb}{44}{8759}{1991}.
\end{thebibliography}
